\documentclass[sigconf]{aamas}

\usepackage{balance} 

\usepackage{amssymb}
\usepackage{amsmath}
\usepackage{amsthm}
\usepackage{booktabs}
\usepackage{enumitem}
\usepackage{graphicx}
\usepackage{color}
\usepackage{url}
\usepackage[utf8]{inputenc}
\usepackage{algorithm}
\usepackage{algorithmic}
\usepackage{tikz}
\usepackage{pgfplots}
\usepackage{mathtools}
\usepackage{subfigure}
\usepackage{xcolor}
\usepackage{placeins}
\newtheorem{assumption}{Assumption}

\doi{FXWO5737}

\makeatletter
\gdef\@copyrightpermission{
  \begin{minipage}{0.2\columnwidth}
   \href{https://creativecommons.org/licenses/by/4.0/}{\includegraphics[width=0.90\textwidth]{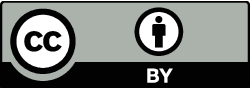}}
  \end{minipage}\hfill
  \begin{minipage}{0.8\columnwidth}
   \href{https://creativecommons.org/licenses/by/4.0/}{This work is licensed under a Creative Commons Attribution International 4.0 License.}
  \end{minipage}
  \vspace{5pt}
}
\makeatother

\setcopyright{ifaamas}
\acmConference[AAMAS '26]{Proc.\@ of the 25th International Conference
on Autonomous Agents and Multiagent Systems (AAMAS 2026)}{May 25 -- 29, 2026}
{Paphos, Cyprus}{C.~Amato, L.~Dennis, V.~Mascardi, J.~Thangarajah (eds.)}
\copyrightyear{2026}
\acmYear{2026}
\acmDOI{}
\acmPrice{}
\acmISBN{}

\renewcommand\footnotetextcopyrightpermission[1]{} 
\title[Decentralized Value Systems Agreements]{Decentralized Value Systems Agreements}

\author{Arturo Hernández-Sánchez}
\affiliation{
  \institution{VRAIN, Universitat Politècnica de València}
  \city{València}
  \country{Spain}}
\email{arthersan@upv.es}
\author{Natalia Criado}
\affiliation{
  \institution{VRAIN, Universitat Politècnica de València}
  \city{València}
  \country{Spain}}
\email{ncriado@upv.es}
\author{Stella Heras}
\affiliation{
  \institution{VRAIN, Universitat Politècnica de València}
  \city{València}
  \country{Spain}}
\email{stehebar@upv.es}
\author{Miguel Rebollo}
\affiliation{
  \institution{VRAIN, Universitat Politècnica de València}
  \city{València}
  \country{Spain}}
\email{mrebollo@upv.es}
\author{Jose Such}
\affiliation{
  \institution{INGENIO (CSIC-Universitat Politècnica de València)}
  \city{València}
  \country{Spain}}
\email{jose.such@csic.es}

\begin{abstract}
One of the biggest challenges of value-based decision-making is dealing with the subjective nature of values. The relative importance of a value for a particular decision varies between individuals, and people may also have different interpretations of what aligning with a value means in a given situation. While members of a society are likely to share a set of principles or values, their value systems—that is, how they interpret these values and the relative importance they give to them—have been found to differ significantly. This work proposes a novel method for aggregating value systems, generating distinct value agreements that accommodate the inherent differences within these systems. Unlike existing work, which focuses on finding a single value agreement, the proposed approach may be more suitable for a realistic and heterogeneous society. In our solution, the agents indicate their value systems and the extent to which they are willing to concede. Then, a set of agreements is found, taking a decentralized optimization approach. Our work has been applied to identify value agreements in two real-world scenarios using data from a Participatory Value Evaluation process and a European Value Survey. These case studies illustrate the different aggregations that can be obtained with our method and compare them with those obtained using existing value system aggregation techniques. In both cases, the results showed a substantial improvement in individual utilities compared to existing alternatives.
\end{abstract}


\newcommand{\BibTeX}{\rm B\kern-.05em{\sc i\kern-.025em b}\kern-.08em\TeX}

\pgfplotsset{compat=1.18}
\begin{document}


\pagestyle{fancy}
\fancyhead{}



\maketitle 


\section{Introduction}
In recent years, there has been an increase in the use of AI systems within decision-making processes in the execution of various tasks, making it essential to ensure that their behavior aligns with human values. The Asilomar AI principles~\cite{asilomarai} emphasize that autonomous systems must respect cultural diversity, human dignity, rights, and freedom. One approach to achieve such alignment is value inference~\cite{valueinference}, which is a process that consists of three steps: first, identifying the values relevant to a decision context from behavioral data; second, estimating the subjective preferences and interpretations individuals assign to these values, which constitutes a personal value system; and third, aggregating these individual systems to infer a common set of societal value preferences and interpretations.

This paper focuses on the third step: the aggregation of value systems. To the best of our knowledge, only Lera et al. \cite{lera2022towards,lera2024aggregating} have addressed this challenge. However, their method is limited to constructing a single consensual value system for the entire population. According to \cite{valueinference}, one of the main challenges in the aggregation of value systems is that reaching a unanimous consensus is often neither possible nor desirable, particularly as individuals may be \emph{``naturally clustered around different consensuses instead of a single consensus that might not be representative of any individual''}. Additionally, some individuals may choose not to cooperate in the consensus process or may wish to reject the final agreement. 
While individuals could in principle retain their personal value systems or engage in negotiation, such approaches scale poorly and often fail when the population is large or when there are substantial differences among agents. 
Another feature of the methods proposed by \citeauthor{lera2022towards} is their centralized nature. In contrast, many AI systems are distributed across multiple AI agents. Decentralized approaches are known to allow for a more participatory, democratic, and self-organizing approach to decision-making processes \cite{malone2004future}. Moreover, decentralized methods are particularly beneficial in sensitive AI applications, such as those in defense or medical decision-making, where privacy and security are critical. Unlike centralized approaches that store sensitive data on a single server—making it vulnerable to cyberattacks—decentralized approaches distribute data across the network and share it only with local neighbors when necessary \cite{nedic2009distributed}.

This paper proposes the first method for the decentralized construction of multiple value systems, where individuals are grouped around them, in a way that each agreed-upon value system aggregates their individual value systems. That is, our approach is able to, in a decentralized manner, create groups of individuals as it aggregates their value systems and have, as an output, multiple agreed-upon value systems. The decentralized network formation process is based on a homophily-based process, as individuals tend to form connections and make common decisions more easily with those who share similar opinions, a phenomenon well-understood in the social sciences \cite{korbel2023homophily,murase2019structural,BERRY2018118}. This approach does not enforce permanent segregation, since agents can continually discover and connect to new neighbors, and links are maintained only when agreements remain within confidence bounds. 
Our main contributions are the following:
\begin{itemize}[noitemsep, topsep=0pt, leftmargin=*]
    \item We propose a novel and decentralized method for aggregating value systems. 
    \item This novel method allows the establishment of multiple agreed-upon value systems around which agents can be grouped to better align with their individual value systems. 
    \item We show the effectiveness of our method in two real-world scenarios based on the Participatory Value Evaluation (PVE) method \cite{mouter2019introduction} and the European Value Study (EVS) \cite{EVS2021}. 
\end{itemize}
\section{Related Work}
Several works have contributed to different steps of the value inference process. For value identification, \cite{liscio2021axies} infer context‑specific values from users’ opinions, while \cite{kiesel2022identifying} infers implicit human values underlying natural‑language arguments. For value system estimation, \cite{siebert2022estimating} estimates individual value systems from survey‑based choices and motivations. The value systems used in our PVE case study were taken from the estimations provided in \cite{siebert2022estimating}.
Regarding the aggregation of value systems — the focus of this paper — some proposals have considered different individual value systems when making decisions. For example, \cite{ajmeri2020elessar} allows one to make decisions considering the value preferences of other individuals involved in the decision-making context. Similarly, \cite{mosca2021elvira} resolves multiuser privacy conflicts in online social networks by integrating the value preferences of affected users. In both cases, decisions are made individually, but they take others’ value preferences into account. Our goal in this step differs: we aim to aggregate multiple value systems into group‑level societal value systems to support value‑based group decision‑making. Beyond value systems aggregation, our work also connects to broader research on moral preference aggregation~\cite{Adler_2016} and pluralistic alignment~\cite{sorensen2024roadmap}, although these approaches do not provide mechanisms for computing multiple group‑specific agreements.
Among the few works that explicitly address the aggregation of value systems is \cite{lera2022towards}, where the authors proposed a method to aggregate them into a single societal value system based on the optimization of a family of distance functions, which they presented as an $\ell_p$-regression problem, with $p$ being a parameter that represents an ethical principle, which ranges from utilitarian (maximum utility, $p=1$) to egalitarian (maximum fairness, $p=\infty$).  While flexible, since the method provides the decision-makers with various single-agreement solution, it does not support the formation of multiple consensual value systems or allow agents to reject the resulting agreement — a limitation highlighted in \cite{valueinference,city35487}. When individuals hold divergent value preferences or interpretations, a single consensus may fail to represent them adequately.
\section{Preliminaries}
\label{sec:Background}
The classical setting of a decentralized optimization model considers a set of $n$ agents $Ag$ arranged into a network defined by a connected graph $G=\left(Ag, E\right)$, where $E$ denotes its set of edges. Each $i\in Ag$ has a private convex objective function $f_i: \mathbb{R}^m \to \mathbb{R}$, $m \in \mathbb{N}$, which is not shared with the other agents in the network. The collective goal is to minimize the global objective function $f: \mathbb{R}^m \to \mathbb{R}$ defined as $f(x)=\sum_{i=1}^{n}f_i(x)$. A widely used method to solve this problem is the Decentralized Gradient Descent (DGD) \cite{nedic2009distributed}, which is defined by the sequences $x_i: \mathbb{N} \to \mathbb{R}^m$ for each $i \in Ag$ as:
\begin{equation}
    x_i\left(t+1\right) = \sum_{j=1}^n a_{ij}x_j\left(t\right) - \alpha\left(t\right)\nabla f_i\left(x_i\left(t\right)\right)
    \label{model_org}
\end{equation}
where $a_{ij}$ denotes the elements of a doubly stochastic matrix $A\in \mathcal{M}_n\left(\mathbb{R}\right)$ --- i.e., the sum of each of its columns and each of its rows is equal to 1 --- called mixing matrix, which determines the topology of the network given by the graph $G$ and the possible weights that could be given to each of the links in the network. The sequence $\alpha\left(t\right)$ is the stepsize, which determines the convergence speed and has to be small enough to ensure the convergence of (\refeq{model_org}) \cite{shi2015extra} and a good accuracy in minimizing the function $f$ \cite{xin2020general}.  For our aggregation problem, we are interested in obtaining consensual solutions, meaning that the sequences defined by (\refeq{model_org}) converge to the same point for all 
 $i\in Ag$, as this will facilitate agents in making collective decisions. The choice of stepsize plays a critical role in obtaining these solutions:  
 a stepsize that tends to zero is known to lead to a consensual solution 
 \cite{bertsekas2003convex,chen2012fast,nedic2014distributed}. 
 Furthermore, as in our case, when we need the solution to be bounded, 
 we must also consider some constraints for the optimization problem we aim to address. In this case, first-order algorithms \cite{sundhar2010distributed,nedic2010constrained} 
 project the solutions obtained in each iteration into a compact and convex feasible set $\mathcal{X} \subset \mathbb{R}^m$. Therefore, the optimization problem they solve is:
\begin{equation}
\begin{array}{c}
    \min \sum_{i=1}^n f_i\left(x\right) \\
    \text{s.t. } x \in \mathcal{X}
\end{array}
\label{opt_problem_projected}
\end{equation}

The solution to this problem can be obtained by the sequences $x_i: \mathbb{N} \to \mathbb{R}^m$, which are defined for each $i \in Ag$ as \cite{nedic2010constrained}:
\begin{equation}
\begin{array}{c}
    x_i\left(t+1\right) = P_{\mathcal{X}}\left[\sum_{j=1}^n a_{ij}x_j\left(t\right) - \alpha\left(t\right)\nabla f_i\left(x_i\left(t\right)\right)\right] \\
    x_i\left(0\right) \in \mathcal{X}
\end{array}
\label{projected_model}
\end{equation}
where $P_{\mathcal{X}}$ denotes the Euclidean projection on to the set $\mathcal{X}$. Convergence of (\ref{projected_model}) to an optimal solution of the optimization problem given by~(\ref{opt_problem_projected}) is guaranteed under the following assumptions \cite{nedic2010constrained}:
\begin{assumption}
     There exists a constant $C > 0$ such that for each $i \in \text{Ag}$, it holds that $\|\nabla f_i(x)\| < C$ for all $x \in \mathcal{X}$.
\end{assumption}
\begin{assumption} The stepsize sequence $\alpha(t)$ must satisfy that \\ $\sum_{t=0}^{\infty}\alpha(t) = \infty$ and $\sum_{t=0}^{\infty}\alpha(t)^2< \infty$.  \label{assump:stepsize}
\end{assumption}

\begin{assumption}There exists a scalar $0<\eta<1$ such that for all $i\in\{1,\dots,n\}$: 
\textit{(i)} $a_{ii}\ge \eta$; 
\textit{(ii)} $a_{ij}\ge \eta$ for all $j\in N_i$; 
\textit{(iii)} $a_{ij}=0$ for all $j\notin N_i\cup\{i\}$; 
\textit{(iv)} $\sum_{j=1}^n a_{ij}=1$; 
\textit{(v)} $a_{ij}=a_{ji}$ for all $j\in\{1,\dots,n\}$.\label{assump:mixing}
\end{assumption}
\section{Problem Statement}
\label{sec:ProbStatement}
Before stating the problem, we introduce how individuals' value preferences and interpretations can be mathematically represented. This has been extensively discussed in the literature \cite{gabriel2020artificial,valueinference,liscio2022values,lera2024aggregating,city35487}, which emphasizes the need to include context, since interpretations and preferences often depend on it. Furthermore, a key feature of Value Theory \cite{schwartz2012overview} is that values guide the selection and evaluation of alternatives.
Inspired by classical Multi-Criteria Decision Making (MCDM) \cite{hwang1981multiple}, we define a \emph{value system} to formalize how agents rank alternatives while accounting for: (i) how alternatives are interpreted relative to a set of values, and (ii) the importance assigned to each value.  
\begin{definition}
    A value system is a $4-tuple$, $\mathcal{V}=\left(V,A,X,\Omega\right)$ where $V$ is a set of values, $A$ is a set of alternatives, $X\in \mathcal{M}_{|A| \times |V|}(\mathcal{I})$ is a decision matrix that contains the evaluations given to each of the alternatives in $A$ under each of the values in $V$ within a determined compact interval $\mathcal{I}\subset \mathbb{R}$ and $\Omega \in \Delta^{|V|} = \{(\omega_1, \dots, \omega_{|V|}) \in \left(0,1\right)^{|V|} :  \sum_{j=1}^{|V|}\omega_j = 1\}$, are the weights assigned to each value in $V$ which determine their order of preference. 
\end{definition}
Intuitively, the decision matrix $X$ captures interpretations, establishing a judgement that measures how desirable is an alternative under an specific value (e.g., how ``safe'' is a private car). The weight vector $\Omega$ captures the relative priorities an individual assigns to each value (e.g., how much safety matters overall). This formal separation directly reflects the conceptual framework of Abia Alonso et al. \cite{city35487}, who emphasize that value aggregation must account for the fact that individuals hold personal judgments—divergent interpretations of the same value—in addition to differing preferences.
\begin{example}
\label{example1}
    Let $Ag=\{1,2,3,4\}$ be a group of four friends who would like to decide what is the best way to travel. They agreed that the values that they want to consider to make this decision are the ones in the set $V=\{\text{Power (P)}, \text{Tradition (T)}, \text{Safety (S)}\}$ and the ways to travel are $A=\{\text{Private Car (PC)},$ $\text{Car Sharing (CS)}\}$.
    For each friend $i\in Ag$, we define its decision matrix $X_i \in \mathcal{M}_{2 \times 3}\left([1,7]\right)$ as follows: 
    $$
   X_1 =  \begin{pmatrix}
        7 & 7 & 7 \\
        1 & 1 & 1
    \end{pmatrix}\;\;\; 
    X_2 =  \begin{pmatrix}
        6 & 3 & 3 \\
        2 & 5 & 5
    \end{pmatrix}\;\;\; 
   $$ 
   $$
   X_3 =  \begin{pmatrix}
        2 & 1 & 3 \\
        6 & 7 & 3
    \end{pmatrix} \;\;\; 
   X_4 =  \begin{pmatrix}
        1 & 1 & 1 \\
        7 & 7 & 7
    \end{pmatrix}  
   $$
       These matrices contain the interpretations they gave to each value in $V$ over each of the alternatives in $A$ using a scale ranging from $1$ (worst) to $7$ (best). For instance, the second friend gives a value of $6$ for using PC over P and a value of $2$ for using CS over P, since he does not have the power to choose where to go at any given time when sharing a car. 
        In addition, he is used to carpooling, so he considers it something traditional; thus, he gives a value of $5$ for the use of CS over T and a value of $3$ for the use of PC over T. Finally, he considers that he does not drive well, so he gives a value of $5$ for the use of CS over S and a value of $3$ for the use of PC over S.
       Next, we define for each $i\in Ag$ its values weight vector $\omega_i \in \mathbb{R}^3$:
   $$
  \Omega_1 = \left(0.4,0.4,0.2\right) \;\;\; \Omega_2 = \left(0.2,0.2,0.6\right)
    $$ 
    $$
      \Omega_3 = \left(0.5,0.2,0.3\right) \;\;\; \Omega_4 = \left(0.3,0.3,0.4\right)
    $$
    for instance, the third friend gives the highest weight for the value "power" (P) and the lowest for the value "tradition" (T).
     The value systems from each $i\in Ag$ are given by $\mathcal{V}_i = \left(V,A,X_i,\Omega_i\right)$. 
\end{example}
As we noted earlier, a single consensus value system often fails to represent populations whose value systems differ substantially or whose members are unwilling to change them. Instead, we aim to form groups of agents that can agree on a group-specific value system representative for that group.  Informally, the goal is to partition the agent set into clusters such that, inside each cluster, agents can reach a shared decision matrix and a shared weight vector that every member finds acceptable. This is more than standard clustering: group formation must respect each agent’s confidence bounds so that agents unwilling to move far from their own value system are not forced into an undesired agreement. This problem is formally defined below:
\begin{definition}
    Let $Ag$ be a set of agents arranged into a network defined by a graph 
    $G = (Ag, E)$, where $E$ denotes the set of edges establishing relationships 
    among the agents, a set of values $V$, and a set of alternatives $A$. 
    For each $i \in Ag$, we define a value system 
    $\mathcal{V}_i = \left(V, A, X_i, \Omega_i\right)$ 
    together with individual \emph{confidence bounds} 
    $\gamma^X_i, \gamma^\Omega_i>0$. 
    The \textbf{problem of aggregating the set of value systems} 
    $\left\{\mathcal{V}_i\right\}_{i\in Ag}$ consists in finding a partition 
    $P$ of the set $Ag$ such that each group $P_k \in P$ has a 
    common value system 
    $\mathcal{V}^*_{P_k}=\left(V, A, X^*_{P_k}, \Omega^*_{P_k}\right)$ 
    that is acceptable to all members $i \in P_k$ within their infividual confidence bounds 
    $(\gamma^X_i, \gamma^\Omega_i)$. These bounds constrain how far $X^*_{P_k}$ 
    and $\Omega^*_{P_k}$ may deviate from $X_i$ and $\Omega_i$ in any agreement.
\end{definition}
The confidence bounds implement the Minimal Decision Divergence (MDD) principle \cite{city35487}, which requires decisions made from a consensus value system to remain as close as possible to those obtained from each individual value system. By restricting the group value system to lie within each agent’s bounds, we ensure that the resulting collective decisions diverge minimally from those of the agents' value systems within each group.
\section{Decentralized Value System Agreements}
\label{sec:Method}
To solve the problem of aggregating a set of value systems, we extend well beyond the projected DGD with a \emph{dynamic network formation} mechanism. At each iteration, agents may connect to or disconnect from others depending on whether updates to their value systems remain within their specified confidence bounds. This ensures that only agents with sufficiently similar value systems remain linked, supporting a homophily-based group formation process. The objective function of each agent is given by its utility function. The goal of the optimization is to identify a partition $P$ of the set of agents $Ag$ such that agents within each group in the partition can agree on a common decision matrix and weight vector that maximize the sum of the individual utilities of the agents within each group of the partition $P$. We define two types of utility functions: one for decision matrices and one weight vectors.
\begin{definition}
\label{def:utility_matrix}
Let $\mathcal{V}_i = (V,A,X_i,\Omega_i)$ be agent $i$’s value system, where $X_i\in \mathcal{M}_{|A| \times |V|}(\mathcal{I})$. The \textbf{decision matrix utility function} of agent $i$ is 
$
u^{X_i}: \mathcal{M}_{|A|\times|V|}(\mathcal{I}) \to \mathbb{R},
$ 
defined for a candidate decision matrix $X = (x_{kj})$ as
$$
u^{X_i}(X) \;=\; -\sum_{j=1}^{|V|}\sum_{k=1}^{|A|} \left(\frac{x_{kj} - x_{i,kj}}{1 - \omega_{i,j}}\right)^2,
$$
where $x_{i,kj}$ are the components of the individual decision matrix $X_i$ and $\omega_{i,j}$ are the components of the weight vector $\Omega_i$.
\end{definition}
The utility function $u^{X_i}$ measures how close a candidate group decision matrix $X$ is to agent $i$’s own decision matrix $X_i$. Its maximum is attained at $X=X_i$ (the agent’s original decision matrix). The denominator $1-\omega_{i,j}$ makes dimensions with larger value weights correspond to narrower parabolas, i.e., the agent is more sensitive to changes in the values they care about. Figure~\ref{fig:utility-graphs} in Example~\ref{example2} illustrates how higher value weights produce narrower parabolas and thus a stronger pull toward the agent’s original evaluations.
\\
Similarly, the \textbf{weight vector utility function} is defined as:
\begin{definition}
Let $\mathcal{V}_i = (V,A,X_i,\Omega_i)$ be agent $i$’s value system. The \textbf{weight vector utility function} of agent $i$ is 
$
u^{\Omega_i}: \Delta^{|V|} \to \mathbb{R},
$ 
defined for a candidate weight vector $\Omega=(\omega_j)_{j=1}^{|V|}$ as
$$
u^{\Omega_i}(\omega) \;=\; -\sum_{j=1}^{|V|} \left(\frac{\omega_j - \omega_{i,j}}{1 - \omega_{i,j}}\right)^2,
$$
where $\omega_{i,j}$ are the components of the individual weight vector $\Omega_i$.
\end{definition}
The intuition behind $u^{\Omega_i}$ is analogous to that of $u^{X_i}$. The function is maximized when the candidate weight vector $\Omega$ equals the agent's own $\Omega_i$. The factor $1-\omega_{i,j}$ in the denominator causes coordinates with larger $\omega_{i,j}$ to produce a more rapid drop in utility for the same change in $\omega_j$. In other words, an agent is less willing to move away from their own weight on a value that they already rank highly. 
\begin{example}
\label{example2}
 Returning to the previous example, we consider the evaluations that each of the agents gave for the alternative of travelling by a private car (PC) under the value safety (S) together with the weights that they gave to this value, which corresponds for each $i\in Ag$ to the element $x_{i,13}$ of its decision matrix and to the component $\omega_{i,3}$ of its weight vector:
    $$x_{1,13}=7 \;\;\;\;  \;\;\;\; x_{2,13}=3 \;\;\;\; x_{3,13}=3 \;\;\;\;x_{4,13}=1$$
    $$\omega_{1,3}=0.2 \;\;\;\;  \omega_{2,3}=0.6 \;\;\;\; \omega_{3,3}=0.3 \;\;\;\; \omega_{4,3}=0.4$$
    The decision matrix utility function for each $i\in Ag$ with respect to this dimension is $u^{X}_{i,13}:\left[1,7\right] \to \mathbb{R}$ defined as
    $$ u^{X}_{1,13}\left(x_{13}\right)=-\left(\frac{x_{13}-7}{1-0.2}\right)^2 \;\;\;\; u^{X}_{2,13}\left(x_{13}\right)=-\left(\frac{x_{13}-3}{1-0.6}\right)^2$$
    $$ u^{X}_{3,13}\left(x_{13}\right)=-\left(\frac{x_{13}-3}{1-0.3}\right)^2 \;\;\;\; u^{X}_{4,13}\left(x_{13}\right)=-\left(\frac{x_{13}-1}{1-0.4}\right)^2$$
\begin{figure}
\vspace{-3pt}
  \centering
    \scalebox{0.6}{
    \begin{tikzpicture}
    \begin{axis}[
        axis lines = left,
        xlabel = \(x_{13}\),
        ylabel = {\(u^{X}_{i,13}(x_{13})\)},
        ylabel style={yshift=5pt}, 
        legend pos=south east, 
    ]

    \addplot [
        domain=1:7, 
        samples=50, 
        color=red,
        style=solid,
        mark=diamond, 
        mark options={scale=0.75},
    ]
    {-((x-7)/(1-0.2))^2};
    \addlegendentry{\(u^{X}_{1,13}\)}
    \addplot [
        domain=1:7, 
        samples=50, 
        color=green,
        style=solid,
        mark=x, 
        mark options={scale=0.75},
    ]
    {-((x-3)/(1-0.6))^2};
    \addlegendentry{\(u^{X}_{2,13}\)}
    \addplot [
        domain=1:7, 
        samples=50, 
        color=blue,
        style=solid,
        mark=|, 
        mark options={scale=0.75},
    ]
    {-((x-3)/(1-0.3))^2};
    \addlegendentry{\(u^{X}_{3,13}\)}
    \addplot [
        domain=1:7, 
        samples=50, 
        color=yellow,
        style=solid,
        mark=triangle, 
        mark options={scale=0.75},
    ]
    {-((x-1)/(1-0.4))^2};
    \addlegendentry{\(u^{X}_{4,13}\)}
    \end{axis}
    \end{tikzpicture}
    }
    \vspace{-4pt}
    \caption{Individual utility functions for value safety (S) and alternative Private Car (PC).}
    \label{fig:utility-graphs}
    \Description{Utility functions for four agents evaluating the safety of traveling by private car. Each curve corresponds to one agent and reaches its maximum at that agent’s original evaluation of the safety of travelling by a private car. Agents with higher weights assigned to safety have narrower parabolas, indicating a faster utility decrease when moving away from their preferred evaluation. Agents with lower safety weights have wider parabolas, showing greater tolerance to deviations. The figure illustrates how value weights affect sensitivity to changes in a specific decision-matrix entry.}
\end{figure}
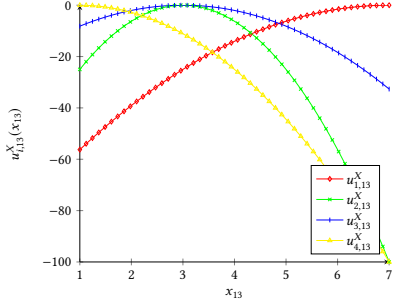 
As shown in Figure \ref{fig:utility-graphs}, for each $i \in Ag$, the point at which the global maximum is reached for the function $u^X_{i,13}$ coincides with the initial evaluation $x_{i,13}$ and the width of each parabola is determined by the weight $\omega_{i,3}$. For instance, in the utility functions from the second, $u^{X}_{2,13}$, and third, $u^{X}_{3,13}$, agents the global maximum is reached in the point with abscissa $3$. However, as the weight assigned to the value S by the second agent is higher than by the third agent, the utility of the second agent decreases faster.
\end{example}
With the individual utility functions defined, we can now formalize how agents reach agreements in a decentralized manner. Consider a set of agents $Ag$, arranged in a network represented by a graph $G=(Ag,E)$, a set of values $V$, and a set of alternatives $A$. Each agent $i \in Ag$ has a value system $\mathcal{V}_i=(V,A,X_i,\Omega_i)$, where $X_i \in \mathcal{M}_{|A|\times|V|}(\mathcal{I})$ and associated confidence bounds $\gamma_i^X, \gamma_i^\Omega > 0$. These bounds specify the tolerance of each agent: how far a candidate group decision matrix $X$ or weight vector $\Omega$ may deviate from their own preferences while still being acceptable.
\\
The initial network edges can be defined in many ways depending on the problem setting and available data. One natural choice is to connect agents whose value systems are sufficiently similar according to their confidence bounds:
\begin{multline}
    E = \{(i,j) \in Ag \times Ag : i\neq j \; \wedge \; d_{|A||V|}(X_i, X_j) \\
    < \min \{\gamma^{X}_i, \gamma^{X}_j\} \; \wedge \; d_{|V|}(\Omega_i, \Omega_j) < \min \{\gamma^{\Omega}_i, \gamma^{\Omega}_j\}\}
    \label{edges_definition}
\end{multline}
where $d_{|A|\times|V|}$ and $d_{|V|}$ are appropriate distance functions on the space of decision matrices $\mathcal{M}_{|A|\times|V|}(\mathcal{I})$ and weight vectors $\Delta^{|V|}$, respectively. Alternative constructions could use physical proximity, social similarity metrics, or Hofstede cultural indices \cite{hofstede2011dimensionalizing}.
\\
For simplicity, we assign equal weights to all edges. A mixing matrix $A \in \mathcal{M}_n(\mathbb{R})$ that assigns equal weights and satisfies Assumption~\ref{assump:mixing} can then be chosen as in \cite{XIAO200465}:
\begin{equation}
    a_{ij} = \begin{cases}
            \varepsilon & \text{if } \{i,j\} \in E \text{ and } i \neq j, \\
            1 - \delta_i \varepsilon & \text{if } i = j, \\
            0 & \text{if } \{i,j\} \notin E \wedge  i \neq j.
         \end{cases}
         \label{mixing_matrix}
\end{equation}
where $\delta_i$ is the degree of node $i$ and 
$
0 < \varepsilon < \frac{1}{\underset{i \in Ag}{\max} \{\delta_i\}}
$. 
With these ingredients, we apply the projected DGD (\ref{projected_model}) to find optimum decision matrices and weight vectors, incorporating our dynamic network formation mechanism, which is captured by the time-varying neighbor sets $N_i(t)$ defined below. For decision matrices, the update for agent $i$ at iteration $t$ is:

\begin{equation}
    \begin{dcases}
        X_i(t+1) = &\mathcal{P}_{\mathcal{M}_{|A|\times|V|}(\mathcal{I})}\bigl[ X_i(t) + \varepsilon \sum_{j \in N_i\left(t\right)} (X_j(t) - \\ & X_i(t)) +
         \alpha\left(t\right) \nabla u^{X}_i(X_i(t))\bigl] \\
        X_i(0)=& X_i 
\end{dcases}
\label{modelo_final_X}
\end{equation}
Similarly, for weight vectors:
\begin{equation}
    \begin{dcases}
        \Omega_i(t+1) = &\mathcal{P}_{\Delta^{|V|}}\bigl[ \Omega_i(t) + \varepsilon \sum_{j \in N_i(t)} (\Omega_j(t) -  \\ 
         &\Omega_i(t)) +\alpha\left(t\right) \nabla u^{\Omega}_i(\Omega_i(t))\bigl] \\
        \Omega_i(0)=& \Omega_i 
\end{dcases}
\label{modelo_final_omega}
\end{equation}
Here, $N_i(t)$ represents agent $i$’s neighbors at iteration $t$, dynamically updated based on both previous neighbors and newly discovered ones:
\begin{equation}
    \begin{cases}       
        N_i(t+1)=& \{j \in N_i(t)\cup N_{\text{new},i}(t): i \neq j \wedge \\ &  d_{|A||V|}(X_i(t), X_j(t))< \min \{\gamma^{X}_i, \gamma^{X}_j\} \wedge \\ & d_{|V|}(\Omega_i(t), \Omega_j(t)) < \min \{\gamma^{\Omega}_i, \gamma^{\Omega}_j\}\} \\
        N_i(0)=& \{j \in Ag : (i,j) \in E\}
\end{cases}
\label{sucesion_vecinos}
\end{equation}
The newly discovered neighbors $N_{\text{new},i}(t)$ may be identified using homophily-based approaches \cite{del2014enhancing,yuan2016efficient}, decentralized network formation algorithms from the social network literature \cite{zhang2022segregation,berenbrink2024asynchronous,jiang2019efficient}, or by assuming that each agent can potentially access all other agents in the network, i.e., $N_{\text{new},i}(t) = Ag \setminus \{i\}$.
Note that the functions $u^{X_i}$ and $u^{\Omega_i}$ are individual \emph{utility} functions that the agent seeks to \emph{maximize}. Accordingly, the update rules in (\ref{modelo_final_X}) and (\ref{modelo_final_omega}) perform \emph{projected gradient ascent} on the agents' utilities: we add the terms $\alpha(t)\nabla u^{X_i}(\cdot)$ and $\alpha(t)\nabla u^{\Omega_i}(\cdot)$ (rather than subtracting them). 
\\
Under the diminishing-stepsize conditions stated in the Preliminaries (Assumption~\ref{assump:stepsize}), and assuming that each final connected component, denoted as $P_k$, of the time-varying communication graph is \emph{jointly connected} over sufficiently long windows, the iterates produced by (\ref{modelo_final_X}) and (\ref{modelo_final_omega}) converge to a common limit for all agents in each component. This common limit is a maximizer of the sum of the agents' utilities restricted to that component:
$$
X^*_{P_k} = \underset{X \in \mathcal{M}_{|A|\times|V|}(\mathcal{I})}{\text{argmax}} \underset{i \in P_k}{\sum} u^{X_i}_i(X) \;\;\;\;\;\; \Omega^*_{P_k} = \underset{\Omega \in \Delta^{|V|}}{\text{argmax}} \sum_{i \in P_k} u^{\Omega_i}(\Omega)
$$
These conclusions follow from results for projected decentralized gradient methods (see e.g. \cite{nedic2010constrained}). The connected components of the limiting graph $G^*$ therefore define the desired partition of $Ag$, and a single global agreement is obtained exactly when $G^*$ is connected.
\section{Case Studies}
\label{sec:CaseStudy}
We evaluate the proposed decentralized aggregation method on two real-world datasets: (i) a Participatory Value Evaluation (PVE) dataset \cite{siebert2022estimating} and (ii) the European Values Study (EVS) \cite{EVS2017}. We take the value systems obtained from these datasets that were previously estimated in \cite{siebert2022estimating} and \cite{lera2022towards}. We compare the results of our method with those obtained by the method proposed in \cite{lera2024aggregating}, under both utilitarian (maximum utility) and egalitarian (maximum fairness) principles, using the same experimental setup to ensure a fair comparison. The experiments were carried out considering the following choices and settings:
\begin{itemize}
    \item \emph{Confidence bounds:}
    We obtained the confidence bounds \((\gamma_i^X,\gamma_i^\Omega)\) from pairwise distances between participants' value systems. Concretely, let
    $
    \mathcal{D}^X = \{d_X(X_i,X_j) : i,j\in Ag,\; i\neq j\}$ and $
    \mathcal{D}^\Omega = \{d_\Omega(\Omega_i,\Omega_j) : i,j\in Ag,\; i\neq j\},
    $
    where we use the  Frobenius distance for decision matrices and the Euclidean distance for weight vectors.
    We then set four representative pairs of bounds:
    $
    \{(\gamma^{X}_k,\gamma^{\Omega}_k)\}_{k=1}^{4}
    = \big\{(Q_1^X,Q_1^\Omega),(Q_2^X,Q_2^\Omega),$ $(Q_3^X,Q_3^\Omega),( \max\mathcal{D}^X,$ $ \max\mathcal{D}^\Omega)\big\}
    $,
    where $Q_k^X$ (resp. $Q_k^\Omega$) is the $k$-th quartile of $\mathcal{D}^X$ (resp. $\mathcal{D}^\Omega$); the fourth pair corresponds to the maximum pairwise distances. These bounds were used to set the initial network configuration, as described in (\ref{edges_definition}).
    
    \item \emph{Neighbor discovery:}
     For simplicity, we assumed that every agent can access the entire network in each iteration, meaning that for each $i \in Ag$, $N_{\text{new},i}(t) = Ag \setminus \{i\}$ for all $t \in \mathbb{N}$.
    
    \item \emph{Distance used in projections:} $\mathcal{P}_{\mathcal{M}_{|A|\times|V|}(\mathcal{I})}$ is taken with respect to the Frobenius norm on matrices, and $\mathcal{P}_{\Delta^{|V|}}$ is taken with respect to the Euclidean norm.

\end{itemize}
\subsection{Participatory Value Evaluation}
\subsubsection{Experimental Setting}
The Participatory Value Evaluation (PVE) \cite{MOUTER202154} is an online system that enables citizens to evaluate government policies. This process allows them to express their preferences and enables governments to select policies that align more closely with them. We use a PVE from 2020~\cite{siebert2022estimating}, which 
aimed to assist the municipality of Sùdwest-Fryslân (Netherlands) in selecting energy policies that align with the preferences of its citizens. Participants were asked to evaluate six energy policies, detailed in Table \ref{tabla_options}, based on their relevance for future decisions. After evaluating these policies, participants were asked to choose the motivations that led them to make these evaluations, from which the implicit values underlying these motivations were determined.
\begin{table}
\centering
 \resizebox{0.45\textwidth}{!}{
\begin{tabular}{cl}
\hline
\textbf{Policy option} & \textbf{Description}                                  \\ \midrule
\(o_1\)                & \textbf{The municipality takes the                             lead and unburdens }\\
                        &  \textbf{you: }The municipality will stay in charge and  \\ 
                        & endorse what residents think is important.\\ 
\(o_2\)                & \textbf{Inhabitants do it themselves:}                          Residents generate  \\
                        & their own energy. \\
\(o_3\)                & \textbf{The market decides:} The                                municipality waits and   \\
                        & sees what the market comes up with. Market players   \\
                        & are obliged to involve the residents in their plans.\\ 

\(o_4\)                & \textbf{Large-scale energy generation                          will take place in a } \\ 
                        & \textbf{small number of places:} This way the municipality   \\
                        & avoids having lots of wind and sun parks. \\
\(o_5\)                & \textbf{Focus on storage}: Súdwest-                             Fryslân will become the \\ 
                        & Netherland’s battery. \\
\(o_6\)                & \textbf{Become the sustainable energy provider                             for the}\\
                        &  \textbf{Netherlands.} \\
                         \bottomrule
\end{tabular}}
\caption{Policy options for Sùdwest-Fryslân PVE \protect\cite{tudelft_energy_sudwest}.}
\label{tabla_options}
\end{table}
 The value systems estimated in \cite{siebert2022estimating} considered a set of five values, listed in Table \ref{tabla_valores}. For each participant, the authors obtained a ranking of these values together with a value-option matrix, which we will treat as the decision matrix of each participant's value system. This matrix determines whether each value is relevant to the corresponding policy option. For our case study, we considered a dataset with the value systems of $n=876$ participants that the authors of \cite{siebert2022estimating} shared with us upon request. This dataset includes the value systems estimated with the method that showed the best performance in their evaluation. We denote the set of all participants as $Ag$. For each $i\in Ag$, its value system is represented as $\mathcal{V}_i=\left(V,A,X_i, \Omega_i\right)$, where: $V$ is a set that contains the values listed in Table \ref{tabla_valores}; $A$ is a set that contains the options shown in Table \ref{tabla_options}; $\Omega_i$ is the weight vector of values from participant $i$, and  $X_i=\left(x_{kj}\right)_i\in \mathcal{M}_{6 \times 5}\left(\left[0,1\right]\right)$ is the decision matrix matrix from participant $i$ whose elements are defined as:
    $$
    x_{kj} = \begin{cases}
        1, & \text{if value } j \text{ is relevant for option } k \\
        0, & \text{otherwise}
        \end{cases}
    $$
\begin{table}
\centering
 \resizebox{0.2\textwidth}{!}{
\begin{tabular}{cl}
\toprule
\textbf{Value} & \textbf{Description}                                                 \\ \midrule
\(v_1\)           & Cost-effectiveness\\
\(v_2\)           & Nature and landscape\\
\(v_3\)           & Leadership  \\
\(v_4\)           & Cooperation\\
\(v_5\)           & Self-determination \\ \bottomrule
\end{tabular}}
\caption{Values inferred from participants' motivations}
\vspace{-4pt}
\label{tabla_valores}
\end{table}
In this context, clustering participants around multiple agreed-upon value systems could help the government better identify the different preferences of citizens, allowing it to propose different policies in each area of the municipality, especially if certain zones have a large majority of people from a specific group. It could also be useful for designing marketing or electoral campaigns targeted at specific groups depending on their preferred policies. For example, individuals in a group that prefers option \(o_2\) would likely have a vision of a self-sufficient, locally protective economy, while those who prefer option \(o_3\) would probably have a global economic perspective. 
\subsubsection{Results}
\label{sec:CaseStudy_utilities}
\begin{figure*}[t]
    \centering
    \subfigure[Agreed-upon decision matrices utilities distribution.]{
        \includegraphics[width=0.45\textwidth]{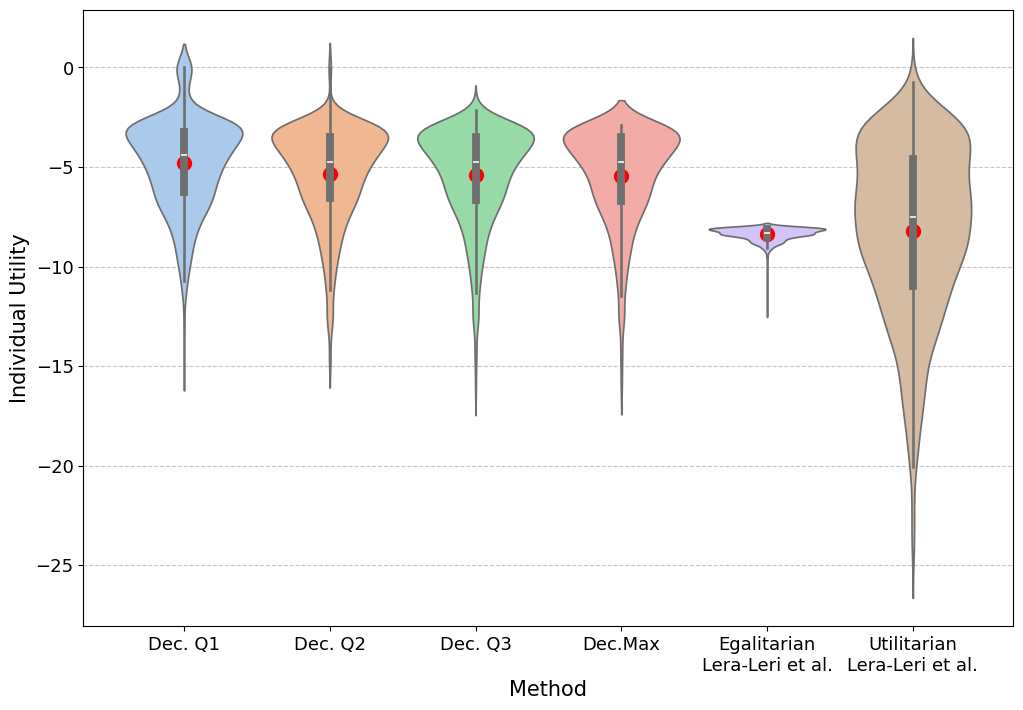}
        \label{fig:matrices_utilities}
        \Description{Distribution of individual utilities for agreed-upon decision matrices in the PVE case study. The figure uses violin plots to show how utilities vary across aggregation methods and confidence bounds. Wider sections indicate higher density of participants with similar utilities. Smaller confidence bounds lead to distributions that are more concentrated toward higher utility values, indicating that more participants are closer to their preferred decision matrices. Mean utilities are highlighted to facilitate comparison across methods.}
    }
    \subfigure[Agreed-upon weight vectors utilities distribution.]{
        \includegraphics[width=0.45\textwidth]{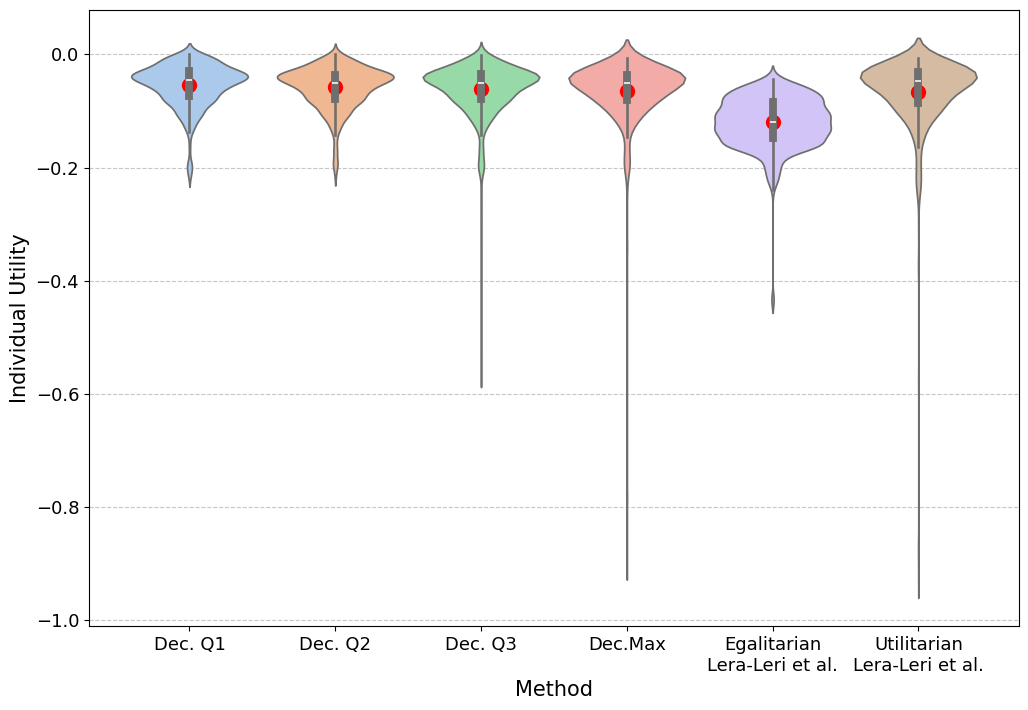}
        \label{fig:vectors_utilities}
        \Description{Distribution of individual utilities for agreed-upon value weight vectors in the Participatory Value Evaluation case study. Violin plots show that tighter confidence bounds produce narrower utility ranges and higher densities near the maximum utility. Compared to centralized aggregation methods, the decentralized approach yields fewer participants with low utilities and more participants with high utilities, indicating improved alignment between group agreements and individual value priorities.
}
    }
    \vspace{-4pt}
    \caption{Utility distributions for the agreed‑upon value systems in the PVE case study. Red points denote mean utilities and the violin widths reflect the density of individual utilities.\\}
    \label{fig:utilities_distributions}
    \Description{}
\end{figure*}
\paragraph{Utilities Distribution} Figure \ref{fig:utilities_distributions} shows the distributions of the individual utilities for the agreed-upon value systems obtained with each aggregation approach. 
As shown in Figure \ref{fig:matrices_utilities}, the utility distributions from the decision matrices, obtained with the different confidence bounds considered in our decentralized model, have similar ranges. However, as the confidence bound is reduced, the density for higher utilities increases. This indicates that we get more participants with higher utilities for the value-option matrices they agreed. The weight vector utility distributions show a more notable improvement as the confidence bound is reduced, as shown in Figure \ref{fig:vectors_utilities}. The range of utilities is considerably shortened by taking the first two quartiles as confidence bounds compared with the other aggregations, obtaining higher utilities with higher density. This means that we get fewer participants with low utilities compared with the other approaches, and we get more participants with the highest utility values.
Regarding the results obtained with the centralized $\ell_p-$\textit{regression} method \cite{lera2022towards}, we got the widest range of utilities considering the utilitarian ethical principle with this method, resulting in very low utilities compared with the other aggregation methods. When aggregating the value-option matrices considering the egalitarian ethical principle, as shown in Figure \ref{fig:matrices_utilities},  a very short range of utilities, from $-12.33$ to $-8$, is obtained, although not as high utilities are obtained for most participants as with the aggregations from our proposed decentralized model. This happens because egalitarian solutions tend to reduce the inequality of utilities between participants, even if it means the utility of most participants is lower. In contrast, utilitarian solutions aim to maximize the utility of most participants, which can lead to increased inequality. The solutions proposed with our decentralized model are based on maximizing the sum of utilities of the agents within a group, so they could be considered an egalitarian solution. However, since the participants may be able to accept or reject the agreement obtained in the group with their confidence bound, it gives more fairness to the solution than the utilitarian one, providing an implicit level of egalitarianism. 
\paragraph{Partition analysis.}
Table \ref{group_desc} shows a description of the partitions obtained using our decentralized approach with the four confidence bounds we considered. As we can observe, as the confidence bounds become smaller, the algorithm produces more groups, and the within-group average distances (both for value–option matrices and for weight vectors) decrease. In other words, smaller confidence bounds make members of the same group to be more similar by splitting the set of participants into more, smaller, and more homogeneous groups. In our experiments, when the smallest bounds were used, the algorithm produced a large principal group together with several much smaller groups (many of them singletons) .
\\
Balancing utility and interpretability, the $Q_2$ confidence bound offers the best trade-off. It achieves nearly the same range and maximum utility values as $Q_1$, but with substantially fewer groups (8 vs. 45; see Table \ref{group_desc}). For this partition, we derived option rankings from the agreed-upon value systems using a standard multi-criteria decision-making (MCDM) method, the Technique for Order of Preference by Similarity to Ideal Solution (TOPSIS) \cite{hwang1981methods}. The resulting rankings, presented in Table \ref{topsis_table}, highlight how different options align with each group’s value system, which vary considerably. For instance, option $o_6$ is ranked as the top choice in group $P_3$ but as the least preferred in group $P_6$.
\begin{table}
\centering
 \resizebox{0.45\textwidth}{!}{
\begin{tabular}{|c|c|c|c|c|}
\hline
\textbf{Confidence Bound} & $Q_1$ & $Q_2$ & $Q_3$ & Max \\
\hline
\hline
\textbf{Number of Groups} & $45$ & $8$ & $3$ & $1$\\
\hline
\textbf{Value-option Matrices Avg. Distance: } & $0.29$ & $0.84$ & $1.86$ & $2.47$\\
\hline
\textbf{Weight Vectors Avg. Distance:} & $0.02$ & $0.08$ & $0.12$ & $0.25$\\
\hline
\end{tabular}}
\caption{Description of the partitions obtained in the PVE case study. Average distances are computed as the mean pairwise distances between group members’ value–option matrices ($\{X_i\}_{i\in Ag}$) and weight vectors ($\{\Omega_i\}_{i\in Ag}$).
}
\label{group_desc}
\end{table}
\begin{table}
\centering
 \resizebox{0.3\textwidth}{!}{
\begin{tabular}{cl}
\toprule
\textbf{Group} & \textbf{Consensual Options Ranking}                \\ \midrule
\(P_1\)          & $o_4 \prec o_6 \prec o_5 \prec o_3 \prec o_1 \prec o_2  $\\
\(P_2\)          & $o_2 \sim o_1 \prec o_4 \prec o_5 \prec o_6 \prec o_3$\\
\(P_3\)          & $o_4 \sim o_1 \sim o_2 \sim o_3 \prec o_5 \prec o_6$  \\
\(P_4\)          & $o_2 \prec o_3 \prec o_5 \sim o_4 \sim o_6 \sim o_1$\\
\(P_5\)          & $o_3 \prec o_4 \prec o_6 \prec o_5 \prec o_1 \sim o_2$\\
\(P_6\)          & $o_6 \prec o_1 \sim o_3 \prec o_2 \prec o_4 \sim o_5$\\
\(P_7\)          & $o_1 \sim o_5 \sim o_4 \prec o_2 \prec o_6 \prec o_3$  \\
\(P_8\)          & $o_2 \prec o_3 \prec o_5 \prec o_4 \sim o_1 \prec o_6$\\
\bottomrule
\end{tabular}}
\caption{Rankings obtained using TOPSIS from the agreed-upon value systems when considering the confidence bound $Q_2$ in the PVE case study.}
\label{topsis_table}

\end{table}
\subsection{European Values Study}
\subsubsection{Experimental Setting}
The \textit{European Values Study} (EVS)~\cite{EVS2021} is a research program on human values.
In our study, we examine the EVS conducted in 2017~\cite{EVS2017}, which includes $36$ participating countries. Specifically, we consider the questions used in the case study described in~\cite{lera2022towards} to compare the results obtained using their aggregation method with ours. In their study, the authors estimated a value system for each country, which they later aggregated. Let $Ag$ denote the set of participating countries. First, they considered the question \textit{Q1 Please say, for each of the following, how important it is in your life}, focusing on the response to the option \textit{v6 Religion}, evaluated on a 1--4 scale, where $1$ indicates that religion is very important in daily life and $4$ indicates that it is not important at all. This evaluation, given by each participant, allowed them to obtain the proportion of individuals that can be considered religious (those rating $1$ or $2$) and non-religious (those rating $3$ or $4$) within each country. Participants considered religious were labeled with the value \textit{religiosity (rl)} and those considered non-religious with the value \textit{permissiveness (pr)}. Each country $i \in Ag$ was then assigned a weight vector $\Omega_i$, consisting of the percentage of religious \textit{(rl)} and permissive \textit{(pr)} participants.
Next, they considered the question \textit{Q44 Please tell me for each of the following whether you think it can always be justified, never be justified, or something in between, using this card}, evaluating different actions on a 1--10 scale, where $1$ means it is never justified and $10$ indicates it is always justified. They specifically considered the response to \textit{v155 Divorce (dv)} and normalized these evaluations within the interval $\left[-1,1\right]$. Additionally, they considered the question \textit{Q27 How would you feel about the following statements? Do you agree or disagree with them?}, which evaluates the statement \textit{v82 Homosexual couples are as good parents as other couples (ad)} on a 1--5 scale, where $1$ signifies strong agreement and $5$ strong disagreement, and they also normalized these results within the interval $\left[-1,1\right]$. The decision matrix $X_i \in \mathcal{M}_{2\times2}\left([-1,1]\right)$ from each country $i \in Ag$ consisted of the evaluations from both religious \textit{(rl)} and permissive \textit{(pr)} participants in each country over the alternatives \textit{(dv)} and \textit{(ad)}.
\\
\begin{figure*}[t]
    \centering
    \subfigure[Agreed-upon decision matrices utilities distribution.]{
        \includegraphics[width=0.45\textwidth]{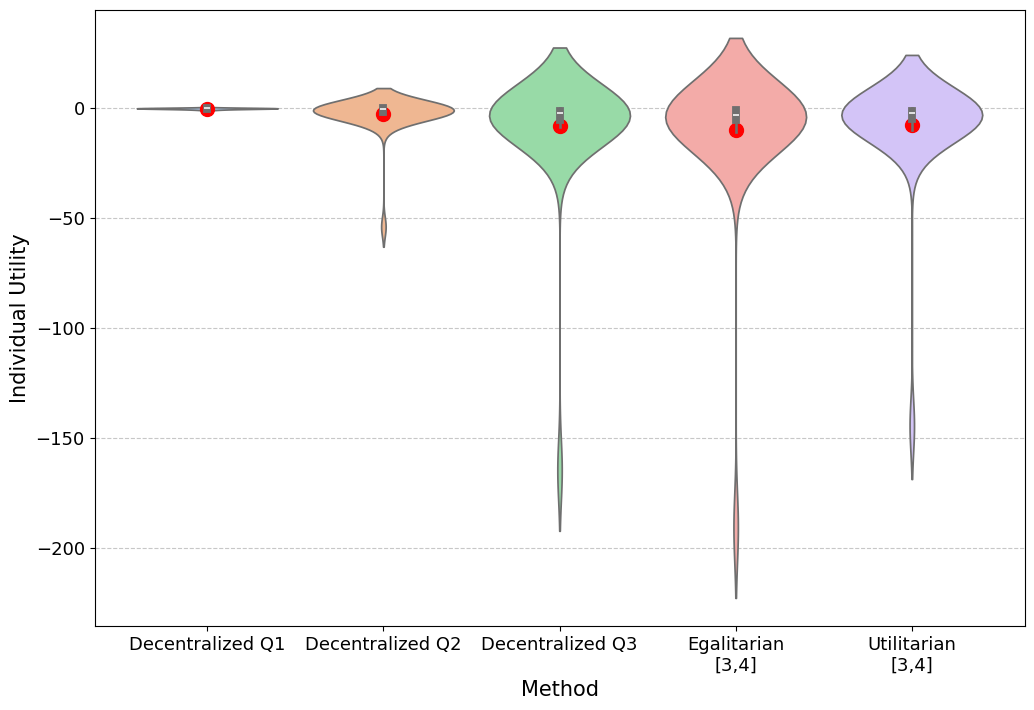}
        \label{fig:matrices_utilities_evs}
        \Description{Distribution of individual utilities for agreed-upon decision matrices in the EVS case study. The violin plots show how utilities change under different confidence bounds. As confidence bounds decrease, the range of utilities becomes smaller.}
    }
    \subfigure[Agreed-upon weight vectors utilities distribution.]{
        \includegraphics[width=0.45\textwidth]{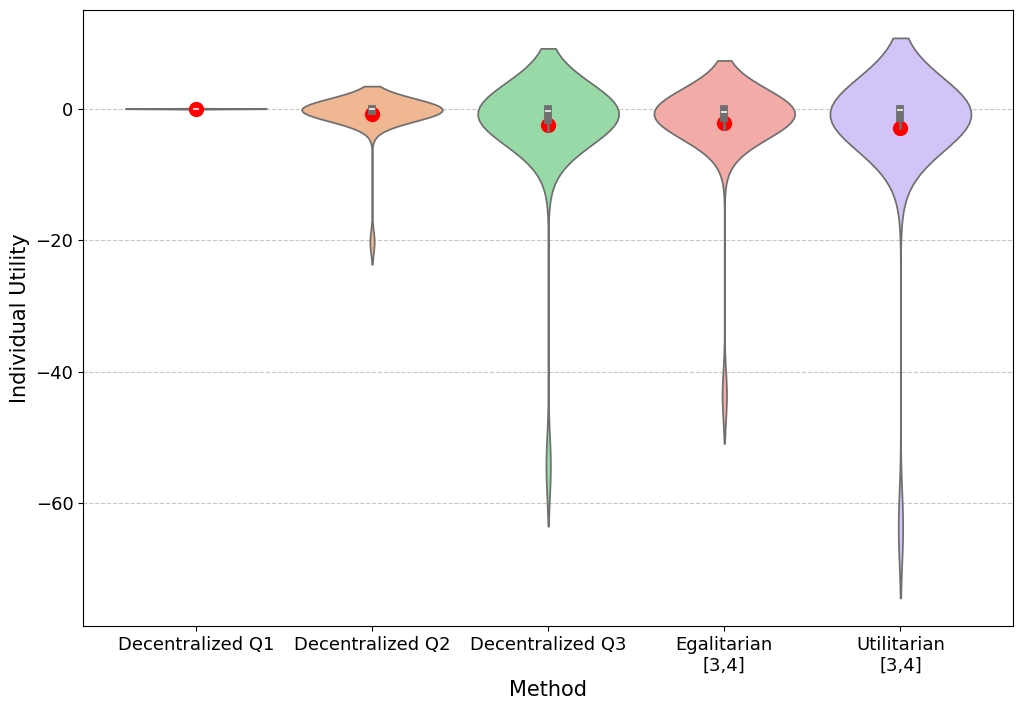}
        \label{fig:vectors_utilities_evs}
    }
    \vspace{-4pt}
    \caption{Distribution of individual utilities for agreed-upon value weight vectors in the EVS case study. Red points denote mean utilities and the violin widths reflect the density of individual utilities.}
    \label{fig:utilities_distributions_evs}
\end{figure*}
In contrast to the PVE case study, where the objective was to guide the selection of alternatives, the EVS case study does not aim to produce rankings of \textit{dv} or \textit{ad}. Instead, our focus is on identifying the communities of countries that emerge under different confidence bounds, analyzing how these communities converge on shared value systems, and comparing the utility distributions across aggregation methods. In this way, the study provides a comparative perspective on value preferences and interpretations across the participating European countries. Here, we omit the largest confidence bound $(\max\mathcal{D}^X, \max\mathcal{D}^\Omega)$ because $(Q_3^X, Q_3^\Omega)$ already yields a single-agreement solution; larger bounds would necessarily produce the same outcome and thus add no new insights. \subsubsection{Results}
\paragraph{Utilities Distribution.} 
The distributions of individual utilities in Figure \ref{fig:utilities_distributions_evs} show that, as in the PVE case study, the range of utilities shrinks as the confidence bound decreases. In this case, however, the distributions for decision matrices and weight vectors display very similar densities, with the decision matrices generally exhibiting a broader range. The range in both distributions decreases significantly when $Q_1$ is used as the confidence bound. This reduction is likely due to the substantial increase in the number of groups formed, which rises from $2$ groups with $Q_2$, to $11$ groups with $Q_1$, as shown in Table \ref{group_desc_evs}. Compared to the baseline in \citeauthor{lera2022towards}, we can see that the range of utilities is significantly reduced when we consider any of the multiple-agreement solutions provided by our method, even with just two agreements obtained with the confidence bound $Q_2$.
\begin{table}
\centering
 \resizebox{0.4\textwidth}{!}{
\begin{tabular}{|c|c|c|c|c|}
\hline
\textbf{Confidence Bound} & $Q_1$ & $Q_2$ & $Q_3$ \\
\hline
\hline
\textbf{Number of Groups} & $11$ & $2$ & $1$ \\
\hline
\textbf{Decision Matrices Avg. Distance: } & $0.10$ & $0.52$ & $0.84$ \\
\hline
\textbf{Weight Vectors Avg. Distance:} & $0.02$ & $0.16$ & $0.33$\\
\hline
\end{tabular}}
\caption{Description of the partitions obtained in the EVS case study. Average distances are computed as the mean pairwise distances between group members’ decision matrices ($\{X_i\}_{i\in Ag}$) and weight vectors ($\{\Omega_i\}_{i\in Ag}$).
}
\label{group_desc_evs}
\end{table}

\begin{figure}
\centering
        \includegraphics[width=0.47\textwidth]{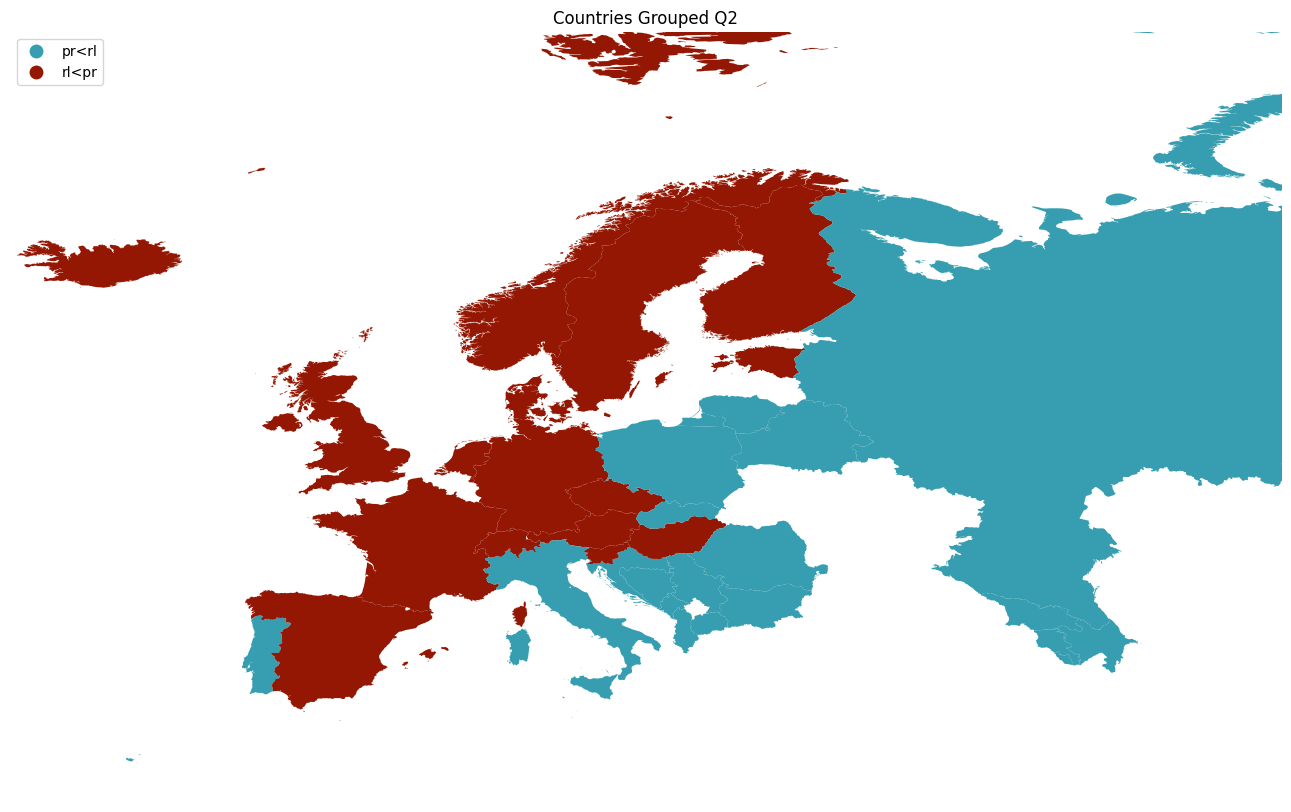}
    \caption{Groups formed when taking $Q_2$ as the global confidence bound in the decentralized value aggregation method.}
    \label{fig:map_q2}
    \Description{Geographical map of Europe showing the groups of countries formed using the decentralized value aggregation method with the second quartile confidence bound. Countries are partitioned into two groups based on their converged value systems. One group, consisting mainly of Eastern European countries together with Italy and Portugal, is more aligned with religiosity, while the other group, composed primarily of Western European countries, is more aligned with permissiveness. The figure visually illustrates how value-based clustering corresponds to regional patterns.}
\end{figure}
\paragraph{Partition Analysis.} 
Table~\ref{group_desc_evs} shows a description of the partitions obtained with each confidence bound. When $Q_2$ is used as the confidence bound, the countries split into two groups (Figure \ref{fig:map_q2}). One group—composed mainly of Eastern European countries together with Italy and Portugal—is more aligned with \textit{religiosity} (\textit{rl}), while the other—primarily Western European countries—is more aligned with permissiveness (\textit{pr}).
Examining the decision matrices agreed upon by each group,
\begin{equation}
   X^*_{\textit{rl}<\textit{pr}} =  \begin{pmatrix}
        -0.62 & -0.38 \\
        -0.37  & 0.15
    \end{pmatrix} \;\;\;\;
    X^*_{\textit{pr}<\textit{rl}} = \begin{pmatrix}
        0.11 & 0.32 \\
        0.23 & 0.54
    \end{pmatrix} \label{dec_matrix}
\end{equation}
we find that, in both cases, the second row—corresponding to the alternative \textit{divorce} (\textit{dv})—receives higher evaluations than \textit{ad}. This indicates that, regardless of whether a group is more aligned with \textit{pr} or with \textit{rl}, divorce tends to be viewed more favorably than same-sex adoption. At the same time, the group more aligned with religiosity (\textit{pr} < \textit{rl}) assigns consistently higher evaluations to both alternatives than the permissiveness-oriented group. 
\section{Conclusions}
\label{sec:conclusions}
In this paper, we addressed several challenges in value system aggregation highlighted in \cite{city35487,valueinference}. We introduced a method that moves beyond enforcing a single consensus—a major limitation of existing approaches—by formulating aggregation as a decentralized optimization problem. This allows groups of agents to converge on value systems that maximize the sum of individual utilities within each group. Through a dynamic communication network, agents naturally self-organize into multiple groups, each forming its own consensual value system. Our case studies show that clustering agents with similar preferences leads to improved utility distributions and better representation of heterogeneous value systems.

For future work, we suggest several directions. First, exploring the effect of weighted graphs, where edges reflect different levels of influence or trust among agents. Second, parameterizing the aggregation to accommodate a range of ethical principles beyond the current utilitarian-egalitarian mix would allow greater flexibility and applicability in diverse settings. Finally, studying the performance of the method when considering alternative mechanisms for how agents discover new neighbors in successive iterations, rather than assuming that every agent can access every other agent in the network.


\begin{acks}
This work was supported
by VAE-VADEM TED2021-131295B-C32, funded
by MCIN/AEI/10.13039/501100011033 and the European
Union NextGenerationEU/PRTR, PID2024-158227NB-C33 funded by MICIU/AEI/10.13039/501100011033/ FEDER, UE, and by the Generalitat Valenciana project PROMETEO CIPROM/2023/23.
\end{acks}

\balance
\bibliographystyle{ACM-Reference-Format} 
\bibliography{VAE_AAMAS26}


\end{document}